\documentclass[preprint,showpacs,showkeys,amsmath,amssymb,prd]{revtex4-1}
\usepackage{epsf}
\usepackage{dcolumn}
\usepackage{amsmath}
\usepackage{amssymb}
\usepackage{color}
\usepackage{bm}
\everymath{\displaystyle}
\usepackage{graphicx}  

\usepackage{amsfonts}

\begin{document}

%
\title{Berry phases in an electric-dipole-moment experiment in an all-electric storage ring}

\author{Alexander J. Silenko\footnote{Email: alsilenko@mail.ru}}
\affiliation{Research Institute for Nuclear Problems, Belarusian State University, Minsk 220030, Belarus\\
Bogoliubov Laboratory of Theoretical Physics, Joint Institute for Nuclear Research,
Dubna 141980, Russia}

\begin{abstract}
Systematic effects caused by the Berry (geometric) phases in an electric-dipole-moment experiment in an all-electric storage ring are considered. We analyze the experimental setup when the spin is frozen and local longitudinal
and vertical electric fields alternate. Due to the Berry phases, the spin rotates about the radial axis. The corresponding systematic error is rather important while it can be canceled with clockwise and counterclockwise beams.
The Berry phases also lead to the spin rotation about the radial axis. This effect can be canceled with clockwise and counterclockwise beams as well. The sign of the azimuthal component of the angular velocity of the spin precession depends on the starting point where the spin orientation is perfect. The radial component of this quantity keeps its value and sign for
each starting point. When the longitudinal and
vertical electric fields are joined in the same
sections without any alternation, the systematic
error due to the geometric phases does not
appear. However, another systematic effect of
the spin rotation about the azimuthal axis takes place and
has opposite signs for clockwise and counterclockwise beams.
\end{abstract}


\maketitle

\section{Introduction}

The geometric phase, Pancharatnam--Berry phase (named after S. Pancharatnam \cite{Pancharatnam} and M. Berry \cite{Berry}) or most commonly Berry phase plays an important role in classical and quantum mechanics. It is a phase difference acquired over the course of a cycle, when a system is subjected to cyclic adiabatic processes, which results from the geometrical properties of the parameter space of the Hamiltonian.

The assumption that the action of the perturbing fields is much weaker than that of the main field means the adiabatic approximation, wherein the time scale over which a time-dependent Hamiltonian varies is long compared to typical
quantum-mechanical oscillation periods. In this case, a quantum-mechanical system remains in the same eigenstate but develops a dynamical phase factor being the geometric (Berry) phase.

The Berry phases lead to rather important effects in spin physics. In particular, the Berry phases manifest themselves in
the muon \emph{g}--2 experiment \cite{PRDfinal}. In this experiment, the motional magnetic field in the muon rest frame caused by electric focusing leads to a change of the \emph{g}--2 frequency (so-called pitch correction \cite{GrangerandFord,FPL,FF,RPJSTAB}). Taking into account the Berry phases is necessary for any electric-dipole-moment (EDM)  experiments. The importance of these phases is mostly caused by their potential to mimic the EDM. They condition the most important systematic error in the neutron EDM experiment \cite{neutronEDM}.  There are also some manifestations of the Berry phases in storage-ring EDM experiments.
The Berry phases resulting in the false EDM effect can originate from the noncommutativity of spin rotations about different axes \cite{proposal}. In three dimensions, when the spin rotates first about one axis and then about a second one, and then the rotations repeat with opposite sign, the total spin rotation is equal to the product of the two rotation angles with respect to the third axis \cite{proposal,protonproposal}. For an EDM experiment in an all-electric storage ring, this situation takes place for the false EDM effect conditioned by a vertical and a longitudinal component of a magnetic field. However, the corresponding systematic error \cite{proposal,protonproposal,YkSBField,SelcukHaciomeroglu} is small due to magnetic shielding.

We consider  another effect taking place in the proton EDM
experiment in an all-electric storage ring and conditioned by the
geometric phase. It appears owing to the noncommutativity of
impacts of the vertical and longitudinal electric fields.

\section{Spin behavior in an all-electric storage ring}

A spin behavior in an all-electric storage ring is defined by the Thomas-Bargmann-Michel-Telegdi equation \cite{Thomas-BMT} generalized to the case of a particle with the EDM \cite{NKFS,RPJ,PRDspin,GBMT,PhysScr}. The main distinguishing feature of the all-electric storage ring for the proton EDM experiment is the use of radial electric field electrodes for bending, quadrupoles for focusing, and sextupoles for a chromaticity and a long spin coherence time \cite{proposal}.
When the magnetic field is absent, the spin motion equation in the Frenet-Serret (FS) coordinate system (see Ref. \cite{JINRLettCylr} for details) reduces to
\begin{equation}\begin{array}{c}
\frac{d\bm\zeta}{dt}=\bm\Omega\times\bm\zeta,\\ \bm\Omega=\frac{e}{mc}\left\{\left(G-\frac{1}{\gamma^2-1}\right)\left(\bm\beta\times\bm
E\right)\right.\\\left.- \frac{\eta}{2}\left[\bm
E-\frac{\gamma}{\gamma+1}\bm\beta(\bm\beta\cdot\bm
E)\right]\right\},
\end{array}\label{eq7}\end{equation}
where $G=(g-2)/2,~\bm\beta=\bm v/c,~\eta=4dm/e,~d$ is the EDM. When the FS coordinate system is used, one describes the momentum evolution in the cylindrical coordinate system and the spin motion relative to the momentum direction. Except for field misalignments and imperfections, the electric field is orthogonal to $\bm\beta$. Thus, we can neglect the term proportional to  $\bm\beta\cdot\bm E$. When the particle momentum in the all-electric EDM storage ring satisfies the condition
\begin{equation}
p=mc\sqrt{\gamma^2-1} =\frac{mc}{\sqrt G},
\label{eq2}\end{equation}
the spin in frozen. In this case, the spin follows the momentum even if the
beam is deviated from the ideal orbit. When the condition (\ref{eq2}) is exactly
satisfied, systematic errors are vanished. We should mention that the
cylindrical coordinates are usually more convenient for a description of the
spin dynamics than the FS ones  \cite{JINRLettCylr}. However, the latter coordinates define
the spin motion relative to the momentum direction. Since this motion is the
only reason for an appearance of systematic errors in the considered case,
the use of the FS coordinates is preferable. In the cylindrical coordinate
system, the beam with the initial azimuthal polarization can acquire a
vertical polarization component even if Eq. (\ref{eq2}) is exact. As a result, the
description of a spin evolution in this system may lead to a consideration
of some fictitious sources of systematic errors.

While the effect of the geometric phase owing to the vertical and longitudinal magnetic fields is rather small, the residual \emph{radial} magnetic field can lead to a significant false EDM signal. This field produces a vertical spin rotation signal which cannot be canceled with clockwise (CW) and counterclockwise (CCW) beams. In addition, the radial magnetic field causes
a vertical separation of two counter-rotating beams \cite{protonproposal}.

It has been mentioned in Ref. \cite{protonproposal} that systematic errors in the proton EDM experiment can be conditioned by the vertical electric field and deviations of the beam momentum from the value (\ref{eq2}). In the present work, we show that these errors can be rather significant and evaluate them. We also determine main properties of spin dynamics caused by these systematic errors. We consider the spin rotation about the radial and longitudinal axes. Unlike precedent investigations, we give a detailed analysis for some special conditions.

In the all-electric storage ring, the longitudinal electric field does not rotate the spin and the main radial electric field rotating the spin about the vertical axis cannot cause a false EDM signal. Therefore, the electric fields cannot create the Berry 
phase from the noncommutativity of spin rotations. The false EDM signal appears owing to an alternation of the vertical and longitudinal electric fields. When the condition (\ref{eq2}) is violated, the first field causes the spin rotation about the radial axis imitating the EDM signal. The second field changing the beam momentum brings a violation of this condition. The resulting effect is caused by misalignments and imperfections of the both fields and these fields do not have a strong impact on the spin. It can be easily checked that the effect considered meets conditions of an appearance of the Berry 
phase.

If the \emph{averaged} beam momentum does not satisfy Eq. (\ref{eq2}), the spin rotation about the
vertical axis appears. Since this rotation can be easily recognized, we
suppose that the beam momentum satisfies Eq. (\ref{eq2}) on the average. To simplify
the analysis, we specify starting points of the beam trajectory where the
beam momentum complies with Eq. (\ref{eq2}) ($\Delta \gamma =0$) and neglect very
small deviations of $\Omega_z$ from zero. When Eq. (\ref{eq2}) is satisfied
even locally, the spin rotates about the radial axis only due to the EDM.

Any spin effect conditioned by the magnetic dipole moment (MDM) and
mimicking the EDM signal may take place only when the condition (\ref{eq2}) is
violated. This situation takes place when the particle momentum is changed
by a longitudinal electric field. We consider second-order systematic errors
(bilinear in small effects caused by the MDM). A violation of the
above-mentioned condition by a vertical electric field and by imperfections
of the main radial electric field can be neglected because $p_r^2$ and $p_z^2$ are small quantities of the second
order. Certainly, the average longitudinal electric field is equal to zero
because \emph{perpetuum mobile} cannot exist. However, this field can be nonzero in some sections of
the ring.

\section{Spin rotation mimicking the EDM signal}

It can be easily shown that the angular velocity of the spin motion in the all-electric storage ring with small perturbations of the azimuthal particle momentum is given by
\begin{equation}
\label{eqOmega}
{\bm \Omega }=\frac{2e}{mc}G^2\gamma \left( \bm\beta \times
\bm E \right)\Delta \gamma -\frac{e\eta }{2mc}\bm E,
\end{equation}
where $\Delta \gamma $ is the deviation from (\ref{eq2}). The main systematic error
consists in local spin rotations about the radial axis and is caused by the
vertical electric field $\bm E^{(v)}=\pm \left|\bm E^{(v)} \right|\bm e_z \equiv \pm E^{(v)}\bm e_z $.

\begin{figure}[htbp]
\begin{center}
\includegraphics[width=5.cm]{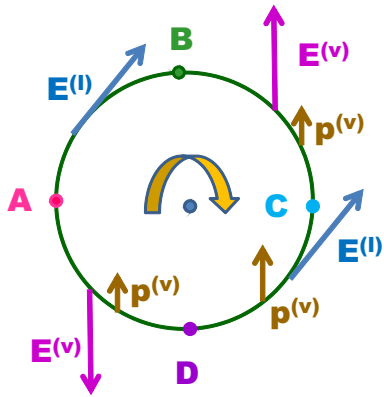} 
\caption{Clockwise beam motion.}
\label{fig1}
\end{center}
\end{figure}

Let us consider the simplest case leading to this systematic error. This
case is shown in Figs. 1, 2. The main electric field is always antiparallel
to $\bm e_r $. We denote $\beta =\left|\bm\beta\right|$.
Regions with nonzero $\Delta \gamma $ and $\bm E^{(v)}$ bounded by the points A,
B, C, and D alternate. We will analyze four cases when both the momentum and
the spin are collinear to the azimuthal direction, $\bm e_\phi$, in 
these starting points.

\subsection{Clockwise beam motion (Fig. 1)} 

a) The starting point is A. At the points B and C, $\Delta \gamma >0$. The angular velocity of the spin motion on the path BC (without the EDM effect) is
\begin{equation}
\label{eq4}
{\bm \Omega }_{BC}=-\frac{2e}{mc}G^2\beta\gamma E^{(v)}|\Delta \gamma|\bm e_{\rho}.
\end{equation}
At the points D and A, $\Delta \gamma =0$. On the path CA, ${\bm \Omega }_{CA}=0$. As a result, the average angular velocity is given by
\begin{equation}
\label{eq5}
{\bm \Omega }=\frac14{\bm \Omega }_{BC}=-\alpha\bm e_{\rho},\quad \alpha=\frac{e}{2mc}G^2\beta\gamma E^{(v)}|\Delta \gamma|.
\end{equation}
This is a systematic error imitating the EDM effect.

We should note that the longitudinal (azimuthal) electric field brings also
the radial velocity component, $\bm\beta^{(r)}$. However, the vector
product $\bm\beta^{(r)}\times
\bm E$ with allowance for the factor $\Delta \gamma$ leads to a third-order correction to $\bm \Omega$. This
correction can be neglected.

b) The starting point is B. At the points C and D, $\Delta \gamma =0$ and $\Delta \gamma <0$, respectively. The false effect is only on the path DA: ${\bm \Omega }_{DA}=-4\alpha\bm e_{\rho}$ and ${\bm \Omega }_{AB}=0$. As a result, ${\bm \Omega }=-\alpha\bm e_{\rho}$.

c) The starting point is C. At the points D and A, $\Delta \gamma <0$. The false effect is only on the path DA: ${\bm \Omega }_{DA}=-4\alpha\bm e_{\rho}$. At the point B and C, $\Delta \gamma =0$. On the path AC, ${\bm \Omega }_{AC}=0$. As a result, ${\bm \Omega }=-\alpha\bm e_{\rho}$.

d) The starting point is D. At the points A and B, $\Delta \gamma =0$ and $\Delta \gamma <0$, respectively. The false effect is only on the path BC: ${\bm \Omega }_{BC}=-4\alpha\bm e_{\rho}$ and ${\bm \Omega }_{CD}=0$. As a result, ${\bm \Omega }=-\alpha\bm e_{\rho}$.

Therefore, the false effect does not depend on the starting point. This
effect can be calculated. The change of the Lorentz factor is equal to
\begin{equation}
\label{eqDelga}
\Delta \gamma =\frac{\beta\Delta p}{mc}=\frac{eR}{mc}\int^{\phi_2}_{\phi_1}{\bm E\cdot\bm e_\phi d\phi},
\end{equation}
where $R$ is the radius of the beam orbit and the angles $\phi_1$ and
$\phi_2$ define the angular size of the section with a nonzero
longitudinal field. In the considered case,
\begin{equation}
\label{eqseven}\begin{array}{c}
\Delta \gamma_{max} =\frac{\pi eR}{2mc}E^{(l)},\qquad \Delta \gamma=\pm\Delta \gamma_{max},\\
|\bm \Omega |=\alpha=\frac{e}{2mc}G^2\beta\gamma E^{(v)}\Delta \gamma_{max}.
\end{array}
\end{equation}
The Berry phase is defined by the angle of the spin rotation about the radial axis during one beam revolution and is equal to
\begin{equation}\label{Berry}\begin{array}{c}
\varphi^{(r)}=\frac{2\pi \Omega_\rho}{\omega_c}=-\alpha T_c,\qquad T_c=\frac{2\pi}{\omega_c},
\end{array}
\end{equation} where $\omega_c$ and $T_c$ are the cyclotron frequency and the cyclotron period, respectively.

\subsection{Counterclockwise beam motion (Fig. 2)} 

\begin{figure}[h]
\begin{center}
\includegraphics[width=5.cm]{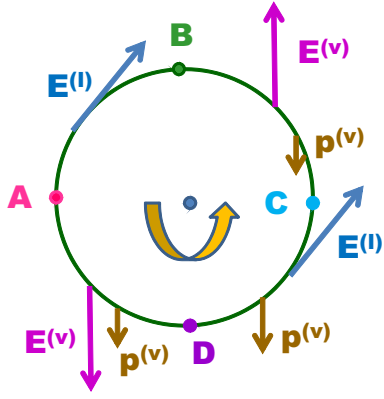} 
\caption{Counterclockwise beam motion.}
\label{fig2}
\end{center}
\end{figure}

a) The starting point is A. At the points D and C, $\Delta \gamma =0$ and $\Delta \gamma <0$, respectively. The false effect is only on the path CB: ${\bm \Omega }_{CB}=4\alpha\bm e_{\rho}$. On the path BA, $\bm \Omega _{BA}=0$.  As a result,
${\bm \Omega }=\alpha\bm e_{\rho}$.

b)  The starting point is B. At the points A and D,  $\Delta \gamma <0$. The false effect is only on the path AD: ${\bm \Omega }_{AD}=4\alpha\bm e_{\rho}$ and $\bm \Omega _{DB}=0$.  As a result, ${\bm \Omega }=\alpha\bm e_{\rho}$.

c)  The starting point is C. At the points B and A, $\Delta \gamma =0$ and $\Delta \gamma <0$, respectively. The false effect is only on the path AD: ${\bm \Omega }_{AD}=4\alpha\bm e_{\rho}$. On the path DC, ${\bm \Omega }_{DC}=0$. As a result, ${\bm \Omega }=\alpha\bm e_{\rho}$.

d)  The starting point is D. At the points C and B,  $\Delta \gamma >0$. The false effect is only on the path CB: ${\bm \Omega }_{CB}=4\alpha\bm e_{\rho}$ and ${\bm \Omega }_{BD}=0$. As a result, ${\bm \Omega }=\alpha\bm e_{\rho}$ and $\varphi^{(r)}=\alpha T_c$.

We can see that the systematic error due to the geometric phases has different signs for the two directions of the beam rotation. The EDM effect remains the same in this case. Thus, the considered systematic error can be canceled with CW and CCW beams.

\section{Spin rotation about the azimuthal direction}

There exists also an effect which does not influence spin dynamics of longitudinally polarized beams in all-electric storage rings. This is the spin rotation about the azimuthal axis which is also conditioned by the longitudinal and vertical electric fields. We suppose that the spin precession about the vertical axis is vanished.

Let us return to Figs. 1, 2. The vertical electric field in the sections BC and DA changes the vertical component of the particle momentum. The maximum value of this component is equal to
\begin{equation}
\label{eq8}
p_{max}^{(v)}=\frac{eR}{c\beta}\int^{\phi_2}_{\phi_1}{E^{(v)}d\phi}.
\end{equation}
In the considered case,
\begin{equation}
\label{eq9}
p_{max}^{(v)}=\frac{\pi eR}{2c\beta}E^{(v)}.
\end{equation}
The vertical component of the particle velocity is given by
\begin{equation}
\label{eq10}
\beta_{max}^{(v)}=\frac{\pi eR}{2mc^2\beta\gamma}E^{(v)}.
\end{equation}
Evidently,
\begin{equation}
\label{eqEvidn}
E=\frac{mc^2\beta^2\gamma}{eR},\qquad \beta_{max}^{(v)}E=\frac{\pi}{2}\beta E^{(v)},
\end{equation}
where $E$ is the main electric field.

We analyze the same cases which have been considered in the precedent section. In the current section, we calculate only the angular velocity of the spin motion acting on the radial spin component.

\subsection{Clockwise beam motion (Fig. 1)} 

The momentum distribution shown at this figure is valid when the starting points are A and B.

a)  The starting point is A. At the points B and C, $\Delta \gamma >0$. The average value of $\beta_z$ on the path BC is equal to $\beta_{max}^{(v)}/2$. The angular velocity of the spin motion on this path reads
\begin{equation}
\label{eq12}
{\bm \Omega }_{BC}=-\frac{e}{mc}G^2\beta_{max}^{(v)}\gamma E\Delta \gamma_{max}\bm e_\phi=-\pi\alpha\bm e_\phi.
\end{equation}
At the points D and A,  $\Delta \gamma =0$. On the path CD, $\beta_{z}=\beta_{max}^{(v)}=const$ and the average value of $\Delta \gamma$  is equal to $\Delta \gamma_{max}/2$. As a result, the angular velocity of the spin motion on this path is given by $\bm \Omega_{CD}=-\pi\alpha\bm e_\phi$. The average angular velocity of the spin motion takes the form
\begin{equation}
\label{eqAvAnV}
\bm \Omega=-\frac\pi2\alpha\bm e_\phi.
\end{equation}
The Berry phase is defined by the angle of the spin rotation about the azimuthal axis during one beam revolution and is given by
\begin{equation}\label{Berryat}\begin{array}{c}
\varphi^{(a)}=\frac{2\pi\Omega_\phi}{\omega_c}=-\frac\pi2\alpha T_c.
\end{array}
\end{equation}

The Berry phase is equal to $\varphi^{(a)}$ if the initial beam polarization is radial or vertical. In the general case, $\bm \Omega=\Omega_\rho \bm e_\rho+\Omega_\phi\bm e_\phi$ and the Berry phase is proportional to the projection of $\bm \Omega$ onto the plane orthogonal to the initial beam polarization. 

b)  The starting point is B. At the points B and C, $\Delta \gamma =0$. On the path BC, $\bm \Omega_{BC}=0$.  The average value of $\Delta \gamma$ on the path CD is equal to $-\Delta \gamma_{max}/2$. The angular velocity of the spin motion on this path reads $\bm \Omega_{CD}=\pi\alpha\bm e_\phi$. On the path DA, the average value of $\beta_z$ is equal to $\beta_{max}^{(v)}/2$ and $\bm \Omega_{DA}=\pi\alpha\bm e_\phi$. As a result, $\bm \Omega=\frac\pi2\alpha\bm e_\phi$ and $\varphi^{(a)}=\frac\pi2\alpha T_c$.

c)  The starting point is C. On the path CD, $\bm \Omega_{CD}=0$. On the path DA, the average value of $\beta_z$ is equal to $-\beta_{max}^{(v)}/2$, $\Delta \gamma=-\Delta \gamma_{max}$, and $\bm \Omega_{DA}=-\pi\alpha\bm e_\phi$. The average value of $\Delta \gamma$ on the path AB is equal to $-\Delta \gamma_{max}/2$. The angular velocity of the spin motion on this path reads $\bm \Omega_{AB}=-\pi\alpha\bm e_\phi$.  On the path BC, $\Delta \gamma=0$ and $\bm\Omega_{BC}=0$. As a result, $\bm \Omega=-\frac\pi2\alpha\bm e_\phi$ and $\varphi^{(a)}=-\frac\pi2\alpha T_c$.

d)  The starting point is D. At the points D and A,  $\Delta \gamma =0$. The false effect is only on the path AC. The average value of $\Delta \gamma $ on the path AB is equal to $\Delta \gamma_{max}/2$. The average value of $\beta_z$ on the path BC reads  $-\beta_{max}^{(v)}/2$. The angular velocity of the spin motion on the paths AB and BC is equal to $\bm\Omega_{AB}=\bm\Omega_{BC}=\pi\alpha\bm e_\phi$.  As a result, $\bm \Omega=\frac\pi2\alpha\bm e_\phi$ and $\varphi^{(a)}=\frac\pi2\alpha T_c$.

\subsection{Counterclockwise beam motion (Fig. 2)} 

The momentum distribution shown at Fig. 2 is valid when the starting points are A and B.

a)  The starting point is A. At the points C and B,  $\Delta \gamma=\Delta \gamma_{max}$. The average value of  $\Delta \gamma$ on the path DC is equal to $\Delta \gamma_{max}/2$. The angular velocity of the spin motion on this path reads  $\bm \Omega_{DC}=\pi\alpha\bm e_\phi$.  On the path CB, $\Delta \gamma=\Delta \gamma_{max}=const$  and the average value of  $\beta_z$ is equal to $-\beta_{max}^{(v)}/2$. As a result, the angular velocity of the spin motion on this path is given by $\bm \Omega_{CB}=\pi\alpha\bm e_\phi$.  The average angular velocity of the spin motion takes the form $\bm \Omega=\frac\pi2\alpha\bm e_\phi$.

b)  The starting point is B. At the points A and D, $\Delta \gamma=-\Delta \gamma_{max}$. On the path AD, the average value of   $\beta_z$ is equal to $-\beta_{max}^{(v)}/2$  and  $\bm \Omega_{AD}=-\pi\alpha\bm e_\phi$. On the path DC, the average value of $\Delta \gamma$ is equal to $-\Delta \gamma_{max}/2$ and $\bm \Omega_{DC}=-\pi\alpha\bm e_\phi$.  As a result, $\bm \Omega=-\frac\pi2\alpha\bm e_\phi$.

c)  The starting point is C. On the path CB, $\bm \Omega_{CB}=0$.  On the path BA, $\beta_z=\beta_{max}^{(v)}=const$, the average value of $\Delta \gamma$ is equal to $-\Delta \gamma_{max}/2$, and $\bm \Omega_{BA}=\pi\alpha\bm e_\phi$. On the path AD, $\Delta \gamma$ is equal to $-\Delta \gamma_{max}$, the average value of $\beta_z$ is equal to $\beta_{max}^{(v)}/2$, and $\bm \Omega_{AD}=\pi\alpha\bm e_\phi$. As a result, $\bm \Omega=\frac\pi2\alpha\bm e_\phi$.

d)  The starting point is D. At the points D and A,  $\Delta \gamma=0$. On the path DC, $\bm \Omega_{DC}=0$. On the path CB,  $\Delta \gamma$ is equal to $\Delta \gamma_{max}$, the average value of $\beta_z$ is equal to $\beta_{max}^{(v)}/2$, and $\bm \Omega_{CB}=-\pi\alpha\bm e_\phi$. On the path BA, $\beta_z=\beta_{max}^{(v)}=const$, the average value of $\Delta \gamma$ is equal to $\Delta \gamma_{max}/2$, and $\bm \Omega_{BA}=-\pi\alpha\bm e_\phi$. As a result, $\bm \Omega=-\frac\pi2\alpha\bm e_\phi$.

The Berry phases are determined similarly to the precedent subsection.

Thus, the longitudinal component of the angular velocity of the spin precession has different signs for the CW and CCW directions of the beam rotation. Therefore, this systematic error can be canceled with CW and CCW beams. Even if the spin precession about the vertical axis is not vanished, this property allows one to eliminate the systematic errors caused by the Berry phases. The azimuthal component of $\bm \Omega$, contrary to the radial one, depends on the starting point where the spin orientation is perfect.

\section{Spin evolution in joined longitudinal and vertical electric fields}

In this section, we consider the spin evolution on condition that the longitudinal and vertical electric fields are joined in the same sections without any alternation. This consideration shows that only an alternation of these fields leads to a systematic error mimicking the EDM signal.

\begin{figure}[h]
\begin{center}
\includegraphics[width=5.cm]{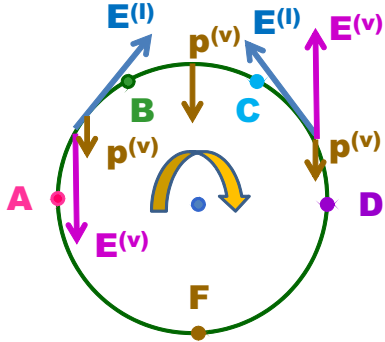} 
\caption{Spin motion in joined longitudinal and vertical electric fields.}
\label{fig3}
\end{center}
\end{figure}

    Let us consider the simple case shown in Fig. 3. In this case, the longitudinal and vertical electric fields are constant into the paths AB and CD and the fields averaged over the ring are equal to zero. The starting point is F. Since $\Delta \gamma=0$ at the points A and D and the beam rotates CW, the average value of  $\Delta \gamma$ on the path AB is given by Eq. (\ref{eqseven}) and is positive. The average value of $\Delta \gamma$ on the path CD is the same. However, the values of $\bm E^{(v)}$ are opposite. As a result, the total spin rotation is equal to zero. The same situation takes place when the beam rotates CCW. Thus, the systematic error due to the Berry phases is absent when the longitudinal and vertical electric fields are joined in the same sections without any alternation.

However, another systematic effect of the spin rotation about the azimuthal axis takes place. Figure 3 shows that the signs of the vertical components of the momentum and velocity are the same in the sections AB and CD. Specifically, we obtain the relations
\begin{equation}\begin{array}{c}
\label{eq14}
\beta_z=-\beta_{max}^{(v)}x, \qquad \Delta \gamma=\Delta \gamma_{max}x, \\ x=\frac{\phi(X)-\phi(A)}{\phi(B)-\phi(A)},
\end{array}\end{equation}
for any point $X\in[A,B]$. As a result, the average angular velocity of the spin motion is given by
$$
{\bm \Omega }_{AB}=\frac{2e}{3mc}G^2\beta_{max}^{(v)}\gamma E\Delta \gamma_{max}\bm e_\phi=\frac23\pi\alpha\bm e_\phi.$$
The average angular velocity of the spin motion on the path CD is the same: $\bm \Omega_{CD}=\bm \Omega_{AB}$. As a result,
\begin{equation}
\label{eq15} \begin{array}{c}
\bm \Omega =\frac{2[\phi(B)-\phi(A)]}{3}\alpha\bm e_\phi,\\ \varphi^{(a)}=\frac{2[\phi(B)-\phi(A)]}{3}\alpha T_c.
\end{array} \end{equation}

When the beam rotates CCW, $\gamma$ remains the same but the vertical components of the momentum and velocity change the signs. In this case,
$$\begin{array}{c}\bm \Omega=-\frac{2[\phi(B)-\phi(A)]}{3}\alpha\bm e_\phi,\\ \varphi^{(a)}=-
\frac{2[\phi(B)-\phi(A)]}{3}\alpha T_c.
\end{array}$$
 Thus, the angular velocity of the spin rotation has opposite directions for the CW and CCW beams.

\section{Discussion and conclusions}\label{CSQ}

The present study shows that local longitudinal and vertical electric fields in the all-electric storage ring may lead to the systematic effects caused by the Berry phases. This takes place when sections with the longitudinal and vertical electric fields alternate. If these fields are perfectly joined, such systematic effects do not appear. Estimates show that the systematic error caused by the Berry 
phases is not small. Nevertheless, it can be canceled with CW and CCW beams. This possibility is conditioned by different signs of the radial component of the angular velocity of the spin precession $\bm \Omega$ for the CW and CCW directions of the beam rotation. The azimuthal component of  $\bm \Omega$ also possesses this property. However, the sign of this component depends on the starting point (where the spin orientation is perfect) while the radial component of $\bm \Omega$ keeps its value for each starting point. When the longitudinal and vertical electric fields are joined in the same sections without any alternation, the systematic error due to the Berry 
phases does not appear. However, another systematic effect of the spin rotation about the azimuthal axis takes place. The angular velocity of the spin rotation has opposite signs for CW and CCW beams.

It is important that both components of $\bm \Omega$ are caused by the same local longitudinal and vertical electric fields. Therefore, the azimuthal component of $\bm \Omega$ can be even helpful because it allows one to evaluate an order of magnitude of the radial component even without the use of the beam with the opposite direction of rotation.

Another important systematic error in the proton EDM experiment in the all-electric storage ring is the residual radial magnetic field.
However, this field can be measured by high-precision magnetometers and the map of magnetic field over the ring can be obtained. In this case, the systematic error due to the radial magnetic field can be transformed into the systematic correction.

\section*{Acknowledgments}

 The work was supported in part by the
Belarusian Republican Foundation for Fundamental Research (Grant No. $\Phi$16D-004) and
by the Heisenberg-Landau program of the German Ministry for Science and Technology
(Bundesministerium f\"{u}r Bildung und Forschung).


\label{last}

\begin{thebibliography}{11}

\bibitem{Pancharatnam}
S. Pancharatnam. Generalized theory of interference, and its applications. Part I. Coherent pencils. Proc. Indian Acad. Sci. \textbf{44}, 247 (1956).

\bibitem{Berry}
 M. Berry. Quantal Phase Factors Accompanying Adiabatic Changes. Proc. Roy. Soc. A \textbf{392}, 45 (1984).

\bibitem{PRDfinal}
G. W. Bennett \emph{et al.} (Muon (\emph{g}--2) Collaboration). Final report of the E821 muon anomalous magnetic moment measurement at BNL.
Phys. Rev. D \textbf{73}, 
072003 (2006).


\bibitem{GrangerandFord}
S. Granger and G.W. Ford. Electron Spin Motion in a Magnetic Mirror Trap. Phys. Rev. Lett. \textbf{28}, 1479 (1972).

\bibitem{FPL}
F. J. M. Farley. Pitch correction in (g-2) experiments. Phys. Lett. B {\bf 42}, 66 (1972); F. J. M.
Farley and E. Picasso. The muon \emph{g}-2 experiments, in {\it Quantum Electrodynamics}, ed. by T.
Kinoshita (World Scientific, Singapore, 1990).

\bibitem{FF} J. H. Field and G. Fiorentini. Corrections to the g - 2 frequency in weak focusing storage devices due to betatron oscillations. Nuovo Cim. A {\bf 21},
297 (1974).

\bibitem{RPJSTAB}
A. J. Silenko. Equation of spin motion in storage rings in the cylindrical coordinate system.
Phys. Rev. ST Accel. Beams \textbf{9}, 
034003 (2006).

\bibitem{neutronEDM}
A. Steyerl, C. Kaufman, G. M\"uller, S. S. Malik, A. M. Desai, R. Golub. Calculation of geometric phases in electric dipole searches with trapped spin-1/2 particles based on direct solution of the Schr\"odinger equation. Phys. Rev. A \textbf{89}, 052129 (2014);
R. Golub, C. Kaufman, G. M\"uller, and A. Steyerl. Geometric phases in electric dipole searches with trapped spin-1/2 particles in general fields and measurement cells of arbitrary shape with smooth or rough walls. Phys. Rev. A \textbf{92}, 062123 (2015).

\bibitem{proposal}
D. Anastassopoulos \emph{et al.} (Storage Ring EDM Collaboration). AGS Proposal: Search for a
permanent electric dipole moment of the deuteron nucleus at the
$10^{-29} e\cdot$ cm level,
https://www.bnl.gov/edm/files/pdf/deuteron$\_$\par\noindent proposal$\_$080423$\_$final.pdf

\bibitem{protonproposal}
D. Anastassopoulos \emph{et al.} (Storage Ring EDM Collaboration). A Proposal to Measure the Proton
Electric Dipole Moment with $10^{-29}~e\cdot$cm
Sensitivity
by the Storage Ring EDM Collaboration,
https://www.bnl.gov/edm/files/pdf/proton$\_$\par\noindent EDM$\_$proposal$\_$20111027$\_$final.pdf

\bibitem{YkSBField}
Y. K. Semertzidis. Geometrical phase from \emph{B}-field in the proton EDM ring.
Internal Storage Ring EDM Collaboration note (2014), unpublished.

\bibitem{SelcukHaciomeroglu}
S. Haciomeroglu. Spin tracking studies for pEDM
experiment. EDM Kickoff meeting @ CERN, https://indico.cern.ch/event/609422/\par\noindent contributions/2463327/attachments/1426326/\par\noindent 2188381/spin$\_$tracking$\_$selcuk.pdf

\bibitem{Thomas-BMT}
L. H. Thomas. The Motion of the Spinning Electron. Nature (London) \textbf{117}, 514 
(1926);
The Kinematics of an Electron with an Axis. Philos. Mag. {\bf 3}, 1 (1927); 
V. Bargmann, L. Michel, and V. L. Telegdi. Precession of the Polarization of Particles Moving in a Homogeneous Electromagnetic Field. Phys. Rev. Lett. {\bf 2}, 435 
(1959). This equation was also derived by J. Frenkel. Die Elektrodynamik des rotierenden Elektrons.
Z. Phys. {\bf 37}, 243 
(1926).

\bibitem{NKFS}
D. F. Nelson, A. A. Schupp, R. W. Pidd and H. R. Crane. Search for an Electric Dipole Moment of the Electron. Phys. Rev. Lett. {\bf 2}, 492 
(1959);
I. B. Khriplovich. Feasibility of search for nuclear electric dipole moments at ion storage rings.
Phys. Lett. B \textbf{444}, 98 
(1998).

\bibitem{RPJ}
A. J. Silenko. Quantum-mechanical description of
the electromagnetic interaction of relativistic particles
with electric and magnetic dipole moments. Russ. Phys. J. {\bf 48}, 748
(2005).

\bibitem{PRDspin} A. J. Silenko. Quantum-mechanical description of spin-1 particles with electric dipole moments.
Phys. Rev. D {\bf 87}, 073015 (2013).

\bibitem{GBMT}
T. Fukuyama and A. J. Silenko. Derivation of Generalized
Thomas-Bargmann-Michel-Telegdi Equation for a Particle with
Electric Dipole Moment. Int. J. Mod. Phys. A \textbf{28}, 1350147
(2013).

\bibitem{PhysScr}
A. J. Silenko. Spin precession of a particle with an electric
dipole moment: contributions from classical electrodynamics and
from the Thomas effect. Phys. Scripta \textbf{90}, 065303 (2015).

\bibitem{JINRLettCylr}
A. J. Silenko. Comparison of spin dynamics in the cylindrical and
Frenet-Serret coordinate systems. Phys. Part. Nucl. Lett.
\textbf{12}, 8 (2015).\\[-34 pt]

\end{thebibliography}
\end{document}